\documentclass[12pt]{article}  
\usepackage{amsmath}
\usepackage{amssymb}
\begin{document}
\title{NONCOMMUTATIVE PARAMETERS OF QUANTUM 
SYMMETRIES AND STAR PRODUCTS}
 \author{P. Kosi\'{n}ski                                                                
                        \\                                                              
           Institute of Physics,                                                        
           University of L\'{o}d\'{z},                                                  
 \\ ul. Pomorska 149/53 90--236                                                    
                L\'{o}d\'{z},   Poland \\                                                   
 email: pkosinsk@krysia.uni.lodz.pl    
\\  \\   J. Lukierski
 \\          Institute of Theoretical Physics,        
 University of Wroc\l aw \\                                                             
   pl. M. Borna 9,                                                                      
           50-205 Wroc\l aw, Poland  \\                                             
     email: lukier@ift.uni.wroc.pl \\                                                  
      and \\                                                                                 
           Department of Theoretical Physics, University of \\              
    Valencia    46100 Burjasot (Valencia), Spain                              
               \\ \\                                                                              
     P.  Ma\'{s}lanka     \\                                                               
    Institute of                                                                           
        Physics, University of L\'{o}d\'{z},                                          
  \\ ul. Pomorska 149/53                                                               
         90--236 L\'{o}d\'{z}, Poland\\                                                 
 email: pmaslan@krysia.uni.lodz.pl    }                                          
                                             
\date{}                                                                                 
\maketitle                                                                              
                                                                                        
\begin{abstract}
The star product technique  translates the framework of local
fields on noncommutative  space--time into nonlocal fields on 
standard
 space--time. We consider the example of fields on $\kappa$--
deformed Minkowski
 space, transforming under $\kappa$--deformed Poincar\'{e} group 
with
 noncommutative parameters. By extending the star product to the
 tensor product of functions on $\kappa$--deformed
 Minkowski space and $\kappa$-deformed
 Poincar\'{e}  group we represent the algebra of noncommutative 
parameters of
 deformed relativistic symmetries
 by functions on classical Poincar\'{e} group.
\end{abstract}

\section{Introduction}
\label{lucsec1}

It has been recognized recently (see. e.g.
\cite{rluk1}--\cite{rluk3})
  that at very short distances, comparable with Planck length
 $\lambda \simeq 10^{-33}$cm, the notion of classical space--time 
manifold should
 be modified. The submicroscopic quantum structure of space--
time implies
 noncommutativity, i.e. one should replace   the classical  
Minkowski coordinates
 $x_{\mu}$ by the generators $\widehat{x}_{\mu}$ of 
noncommutative algebra.
 Assuming the formula (see e.g.
   \cite{rluk4}, \cite{rluk5})$^{[1]}$
\footnotetext[1]{We shall  restrict
 our considerations to the case when the noncommutative algebra 
of space--time
 is generated only by $\widehat{x}_{\mu}$. In general case, as in 
the first
 model of noncommutative space--time by Snyder
  \protect\cite{rluk6}, the operator
 basis of the algebra is extended by  other   operators (see e.g.
 \protect\cite{rluk7}) }

\begin{equation}
\label{eluk1}
\left[  \widehat{x}_{\mu}, \widehat{x}_{\nu} \right]
= \theta_{\mu\nu} (\widehat{x} )
=   \theta_{\mu\nu}^{(0)}  +
\theta_{\mu\nu}^{(1)} {}^{\rho}\widehat{x}_{\rho}
+
\theta_{\mu\nu}^{(2)}  {}^{\rho\tau}\widehat{x}_{\rho}
\widehat{x}_{\tau} + \ldots
\end{equation}
with $\theta_{\mu\nu} (\widehat{x} )$ restricted by Jacobi  
identities, one
  arrives at different  models of noncommutative  space--time   
geometry. The
simplest case is obtained if
$\theta_{\mu\nu} ({x} )$ is a constant
$(\theta_{\mu\nu} ({x} )\equiv \theta_{\mu\nu}^{(0)})$. Such
 a deformation, firstly   advocated by Dopplicher, Fredenhagen and 
Roberts
  \cite{rluk1}  
  has been
  recently extensively studied  in string theory as
 describing world volume coordinates of
 $D$--branes (see e.g. \cite{rluk7}--\cite{rluk9}).
 In such a deformation the relativistic symmetries remain classical, 
what simplifies
greatly  the formalism of corresponding noncommutative theory. 
Indeed, if we
 put $\theta_{\mu\nu} (\widehat{x} )= \theta_{\mu\nu}^{(0)}$ the
 relations (\ref{eluk1}) remain invariant under the shifts
 $\widehat{x}^{\prime}_{\mu} = \widehat{x}_{\mu} + a_{\mu}$ where 
$a_{\mu}$ are
 classical commutative transformations. If the rhs of (\ref{eluk1})
 depends on $\widehat{x}_{\mu}$ the translations preserving  the 
algebraic structure
 of space--time become noncommutative. The simplest framework 
is provided if the
 rhs of (\ref{eluk1}) is linear, i.e.
\begin{equation}
\label{eluk2}
\left[  \widehat{x}_{\mu}, \widehat{x}_{\nu} \right]
=
\theta_{\mu\nu}^{(1)}  {}^{\rho}\widehat{x}_{\rho}
\end{equation}
In such a case the translations $\widehat{x}^{\prime}_{\mu} =
\widehat{x}_{\mu} + \widehat{v}_{\nu}$ commute with  space--time 
algebra
\begin{equation}
\label{eluk3} \left[  \widehat{x}_{\mu}, \widehat{v}_{\nu} \right]
= 0
\end{equation}
and form themselves second copy of the algebra (\ref{eluk2})

\begin{equation}
\label{eluk4}
\left[  \widehat{x}_{\mu}, \widehat{v}_{\nu} \right]
=
\theta_{\mu\nu}^{(1)}  {}^{\rho}\widehat{v}_{\rho}
\end{equation}
 In the general case the translations  form another  copy of algebra
 (\ref{eluk1}), but     the invariance under translations implies 
nontrivial
 braiding relations between the noncommutative algebra of space--
time coordinates
 (\ref{eluk1}) and the translations  algebra  (we use property
 $\theta_{\mu\nu}= - \theta_{\nu\mu}$):
\begin{equation}
\label{eluk5}
\left[  \widehat{x}_{\mu}, \widehat{v}_{\nu} \right]
=
{1\over 2}
 \left\{
 \theta_{\mu\nu}(\widehat{x} +\widehat{v})
 - \theta_{\mu\nu}(\widehat{x}) - \theta_{\mu\nu}
 (\widehat{v}) \right\}
\end{equation}

The noncommutativity of space--time translations implies
necessarily the modification of space--time symmetries. In
particular, one can pose the question  for which functions
 $\theta_{\mu\nu}(\widehat{x})$ in (\ref{eluk1}) the
 noncommutative translations $\widehat{v}_{\mu}$ can be
 extended to the quantum Poincar\'{e} group, describing the Hopf
 algebra of deformed relativistic symmetries. The classification
 of noncommutative translations which can be extended to standard
 (non--braided) quantum Poincar\'{e} group   was given by
 Podle\'{s} and Woronowicz \cite{rluk10}. In particular, if we
 wish
 to maintain the \underline{classical} nonrelativistic
 $O(3)$--symmetries, the choice of the deformation is unique -- one
 obtains the standard form of $\kappa$--deformation of
 relativistic  symmetries (\cite{rluk11}--\cite{rluk13}).

 The aim of  this talk is to describe the $\kappa$--deformed
 field theory in the commutative framework of classical fields,
 with the noncommutative parameters of $\kappa$--deformed
 Poincar\'{e} group described  by  commutative
 parametrization. For that purpose the star  product on
 $\kappa$--deformed Minkowski space     \cite{rluk14}, identical
 to the star product on the subalgebra of noncommutative
 translations, is  extended to ten generators
 ($\widehat{v}_{\mu}, \widehat{\Lambda}_{\ \kappa}^{\nu}$) of
  $\kappa$--deformed Poincar\'{e} group.

  It  appears that  due to the fact that the cross relations  between  
Lorentz
  generators $\widehat{\Lambda}_{\mu\nu}$ and translations
  $\widehat{v}_{\mu}$ are quadratic, our  extended star product
  goes beyond the CBH formula describing star products for
  Lie--algebraic or Lie--superalgebraic structures.

  The plan of our presentation is the following:

  In the Section 2 we describe the $ \kappa$--deformed
  Poincar\'{e}  group and recall the star product for the fields
  defined on $\kappa$--deformed Minkowski space--time. In Sect. 3
  we introduce the  star product for functions on
  $\kappa$--deformed Poincar\'{e} group. In Sect. 4 we present
  final remarks and outlook.

\section{$\kappa$--Deformed Poincar\'{e} Group and Fields on
$\kappa$--Deformed Minkowski Space}

The $\kappa$--deformed Poincar\'{e} group is described  by the
deformed noncommutative group parameters $(\widehat{v}_{\mu},
\widehat{\Lambda}^{\mu}_{\ \nu})$ satisfying the algebraic relations
 \cite{rluk15,rluk16}
\begin{subequations}
\label{eluk21}
\begin{eqnarray}
 \left[  \widehat{v}_{\mu}, \widehat{v}_{\nu}
\right]
& = & { i\over \kappa} \left(
  \delta^{0}_{\ \mu} \,
 \widehat{v}_{\nu} -  \delta^{0}_{\ \nu} \, \widehat{v}_{\mu}
  \right)
\label{eluk21a}
  \\
\left[  \widehat{\Lambda}^{\mu}_{\ \nu}, \widehat{v}_{\rho} \right]
& = &
 - {i \over \kappa}
 \left\{
 \left(
 \Lambda^{\mu}_{\ 0} -
  \delta^{\mu}_{\ 0} \,
\right)
 \widehat{\Lambda}_{\rho\nu} +
\left(
 \widehat{\Lambda}_{0\nu} -
  \eta_{0\nu} \,
\right)  \delta^{\mu}_{\ \rho}\right\}
\label{eluk21b}
 \\
\left[  \widehat{\Lambda}^{\mu}_{\ \nu}, \widehat{\Lambda}^{\rho}_{\ 
\tau}
\right]  &= & 0
 \label{eluk21c}
\end{eqnarray}
\end{subequations}
with the constraints $\Lambda \Lambda^{T} = 
\Lambda^{T}\Lambda = 1$
or
\begin{equation}
\widehat{\Lambda}^{\mu}_{\ \nu}
\,
\widehat{\Lambda}^{\tau}_{\ \nu} =
\widehat{\Lambda}^{\nu}_{\ \mu}
\widehat{\Lambda}^{\nu}_{\ \tau}
= \eta^{\mu\tau} = \eta_{\mu\tau}
\label{eluk22}
\end{equation}
where ${\rm diag}\,  \eta=(1,1,1,-1)$.

The relations (\ref{eluk21}) were firstly obtained \cite{rluk15}
 by  the quantization of Poisson--Lie bracket for the functions on
  Poincar\'{e}  group with the following classical $r$--matrix
\begin{equation}\label{eluk23}
  r = {1\over \kappa} N_{i}\, \Lambda \, P_{i} \, ,
\end{equation}
where $P_{i}$ are three--momenta and $N_{i}\equiv   M_{i0}$ are
 Lorentz boost generators. Another way to obtain the relations
 (\ref{eluk21a})--(\ref{eluk21c}) was to construct
  the dual Hopf algebra to the
 $\kappa$--deformed Poincar\'{e}   algebra
 $\mathcal{U}_{\kappa}(\mathcal{P}_{4})$ written in bicrossproduct
 basis (\cite{rluk13},\cite{rluk16}). The coproduct for
  $\widehat{v}_{\kappa}, \widehat{\Lambda}^{\kappa}_{\ \nu})$
  remain undeformed

\begin{subequations}
\label{eluk24}
\begin{eqnarray}
\Delta(\widehat{v}_{\mu}) & = &
\widehat{v}_{\nu} \otimes
\widehat{\Lambda}^{\ \nu}_{\mu} + \mathbf{1} \otimes
\widehat{v}_{\mu}
\label{eluk24a}
\\
\Delta(\widehat{\Lambda}^{\mu}_{\ \nu} ) & = &
\widehat{\Lambda}^{\mu}_{\ \rho} \otimes
\widehat{\Lambda}^{\rho}_{\ \nu}
\label{eluk24b}
\end{eqnarray}
\end{subequations}
i.e. the composition of two quantum Poincar\'{e} group 
transformations is described by standard classical formulae.

The $\kappa$--deformed Minkowski space described in the formula
 (\ref{eluk24a}) by $\widehat{x}=\widehat{v}_{\mu} \otimes
 \mathbf{1}$ satisfies the relations (\ref{eluk21a}), or, more
 explicitly,
\begin{equation}\label{eluk25}
  \left[ \widehat{x}_{0},
  \widehat{x}_{i}\right] = {i\over \kappa}
  \widehat{x}_{i}
   \qquad
  \left[ \widehat{x}_{i},
  \widehat{x}_{j}\right] = 0
\end{equation}
The $\kappa$--deformed field theory is described by the operator
functions $\Phi_{A}(\widehat{x})$. Following the arguments given
in \cite{rluk17},\cite{rluk14} we shall use for the fields
$\Phi_{A}(\widehat{x})$ the $\kappa$--deformed  Fourier transform
\begin{equation}\label{eluk26}
\Phi_{A}(\widehat{x}) =
{1\over (2\pi)^{4}} \int d^{4} p \,
\widetilde{\Phi}_{\kappa}(p) :
e^{ip\widehat{x}}:
\end{equation}
where
\begin{equation}\label{eluk27}
:e^{ip\widehat{x}}: \, \equiv
e^{- ip_{0}\widehat{x}_{0}}
e^{i\vec{p}\,  \vec{\widehat{x}} }
\end{equation}

and
\begin{equation}\label{eluk28}
\widetilde{\Phi}_{\kappa}(p) =
e^{ {3p_{0} \over \kappa}}\,
\widetilde{\Phi}\left( e^{{p_{0}\over \kappa}} \, \vec{p},p_{0}
\right)
\end{equation}
We have
\begin{equation}\label{eluk29}
:e^{ip\widehat{x}}: \,
:e^{ip^{\prime}\widehat{x}}:
 = :e^{i\Delta^{(2)}(p,p^{\prime})\widehat{x}}:
\end{equation}
where $\Delta^{(2)}_{\mu} = (
\Delta^{(2)}_{0} = p_{0} + p_{0}^{\prime},
\Delta^{(2)}_{i} = p_{i}
e^{{p_{0}^{\prime} \over \kappa}} + p^{\prime}_{i})$.

The algebraic relation (\ref{eluk29}) is translated into star
product framework by the replacement $\widetilde{x}_{\mu} \to
x_{\mu}$ where $x_{\mu}$ are classical  space--time coordinates 
and the ordering in eq.  (\ref{eluk27}) is reflected in explicite
choice of the  star multiplication:
\begin{equation}\label{eluk210}
  e^{ipx} \star e^{ip^{\prime}x} =
e^{i\Delta^{(2)}(p,p^{\prime}){x}}
\end{equation}
i.e. after replacement $\Phi(\widehat{x}) \to \phi(x)$ one gets
\begin{equation}\label{eluk211}
  \phi(x) \star \chi(x) =
  {1\over (2\pi)^{4}} \int d^{4} p \ d^{4}p^{\prime}
  \widetilde{\phi}_{\kappa}(p)
  \widetilde{\chi}_{\kappa}(p^{\prime})
e^{i\Delta^{(2)}(p,p^{\prime}){x}}
\end{equation}

In the following section we shall extend the star product
(\ref{eluk210}--\ref{eluk211}), valid for the noncommutative
translations, to the whole quantum $\kappa$--deformed Poincar\'{e}
group.

\section{The Star Product for $\kappa$--Deformed Poincar\'{e}
Group}

In order to extend the action of Poincar\'{e}  group on Minkowski 
space
 to the noncommutative case we 
have to replace the classical Poincar\'{e} group
 by its $\kappa$--deformed counterpart
\begin{equation}\label{eluk31}
  \left(a_{\mu}, \Lambda^{\mu}_{\ \nu} \right) \Longrightarrow
  \left(\widehat{a}_{\mu}, \widehat{\Lambda}^{\mu}_{\ \nu} \right)
\end{equation}
The noncommutativity of symmetry group parameters raises the
question of the physical interpretation of deformed symmetries. In
this chapter we shall show how one can replace  the operator
algebra of functions on $\kappa$--deformed Poincar\'{e} group
 (\ref{eluk21a}--\ref{eluk21c}) by the functions on classical
  Poincar\'{e} group, with suitably chosen star product
  multiplication.

  We shall    consider the algebra of  the
  following ordered exponentials
\begin{equation}\label{eluk32}
: e^{i(\alpha_{\mu} \widehat{v}^{\mu}
 + b^{\nu}_{\mu}
  \widehat{\Lambda}^{\mu}_{\ \nu} )}: \, =
 e^{-i \alpha_{0}\widehat{v}_{0}}
 e^{i \vec{\alpha}\vec{v} }
 e^{i b_{\nu}^{\mu} \widehat{\Lambda}_{\mu}^{\ \nu}}
 \end{equation}
 The    product of two ordered exponentials (\ref{eluk32}) is
 given by the formula:

\begin{eqnarray}\label{eluk33} 
&&  : e^{i\alpha_{\mu} \widehat{v}^{\mu}  + ib^{\nu}_{\ \mu}
  \widehat{\Lambda}^{\mu}_{\ \nu} }: :
   e^{i\alpha_{\mu}^{\prime} \widehat{v}^{\mu}
 + ib^{\prime \nu}_{\mu}
  \widehat{\Lambda}^{\mu}_{\ \nu} }: \cr\cr
 && =: e^{ i\Delta^{(2)}_{\mu}(\alpha,\alpha^{\prime}) 
\widehat{v}^{\mu}} : e^{i( b^{\nu}_{\ \mu} f^{\mu}_{\nu} 
(g^{\rho}_{\sigma} (
\widehat{\Lambda},\vec{\alpha}^{\prime}), \alpha ) + 
b^{\nu}_{\ \mu} \widehat{\Lambda}^{\mu}_{\ \nu}) } 
\end{eqnarray} 
where
\begin{subequations} \label{eluk34}
\begin{eqnarray} 
  e^{- i\lambda\alpha^{0}} \widehat{\Lambda}^{\kappa}_{\  \nu}
  e^{i \lambda \alpha^{0}}
&  = & f^{\kappa}_{\ \nu} \left(
\widehat{\Lambda},\lambda \right)
\label{eluk34a} 
\\ \cr  e^{- i\vec{\lambda}\vec{\alpha}}
   \widehat{\Lambda}^{\kappa}_{\  \nu}
  e^{i \vec{\lambda} \vec{\alpha}}
&  = & g^{\kappa}_{\ \nu} \left( 
\widehat{\Lambda}, \vec{\lambda} \right) 
\label{eluk34b} 
\end{eqnarray}
\end{subequations}
The functions $f^{\kappa}_{\ \nu}$ and $g^{\kappa}_{\ \nu}$ can be
calculated explicitly. The functions defined by  (\ref{eluk34a}) read:
\begin{eqnarray}\label{eluk35}
  f^{0}_{\ 0}( \widehat{\Lambda}, \lambda )
  &= & \tanh {\lambda \over \kappa}
  \left(
  {1 + \coth {\lambda \over \kappa} \widehat{\Lambda}^{0}_{\ 0}
  \over 1 + 
 \tanh {\lambda \over \kappa} \widehat{\Lambda}^{0}_{\ 0} }
\right)
\cr\cr  f^{0}_{\  k}( \widehat{\Lambda}, \lambda )
  &= & \left( \cosh {\lambda \over \kappa}\right)^{-1}
  \,
  { \widehat{\Lambda}^{0}_{\ k}
  \over 
1 + \tanh {\lambda \over \kappa} \, \widehat{\Lambda}^{0}_{\ 0} }
\cr\cr f^{ k }_{\ 0}( \widehat{\Lambda}, \lambda )
  &= & \left( \cosh {\lambda \over \kappa}\right)^{-1}  \,
  { {\Lambda}^{k}_{\ 0 }
  \over 1 + \tanh {\lambda \over \kappa} \, \widehat{\Lambda}^{0}_{\ 
0} }
\cr\cr f^{i}_{\ k}
 ( \widehat{\Lambda}, \lambda )
  &= &  {
\widehat{\Lambda}^{i}_{\ k} +
\tanh {\lambda 
\over \kappa} \left(
\widehat{\Lambda}^{0}_{\ 0}\, \widehat{\Lambda}^{i}_{\ k} -
\widehat{\Lambda}^{i}_{\ 0}\, \widehat{\Lambda}^{0}_{\ k} \right)  
 \over  1 + \tanh {\lambda \over \kappa}\,
 \widehat{\Lambda}^{0}_{\ 0} }
\end{eqnarray}
The calculation of  (\ref{eluk34b}) is more
complicated, but also possible. 
They are described by the following set of relations
\begin{subequations}  
\begin{eqnarray}      
e^{- i\lambda \, {a}^{k}} \, \Lambda^{0}_{\ 0}\,  e^{ i\lambda \, 
 {a}^{k} }
&= & {\left( {\Lambda^{0}_{\ 0} - 1 \over 2 \kappa^{2} } \right)
\lambda^{2} -
{ \Lambda^{k}_{ \ 0} \over \kappa } \lambda +
 \Lambda^{0}_{ \ 0} \over 
{\left(\Lambda^{0}_{\ 0} -1 \right) \over 2\kappa^{2} }
\,  \lambda^{2} -  
{\Lambda^{k}_{\ 0} \over \kappa}\,  \lambda + 1 }
\label{eluk22a}
\end{eqnarray}
\begin{eqnarray}
 e^{- i\lambda \, {a}^{k}} \, \Lambda^{k}_{\ 0}\,  e^{ i\lambda\, {a}^{k} }
& = & { -                                                                
\left( \Lambda^{0}_{\ 0} - 1 \right) \, { \lambda \over  \kappa }  + 
 \Lambda^{k}_{ \ 0}    
\over  {\left(\Lambda^{0}_{\ 0} -1 \right) \over 2\kappa^{2} } \,  
\lambda^{2} -                                                              
{\Lambda^{k}_{\ 0} \over \kappa}\,  \lambda + 1  }                            
\label{eluk22b} 
\end{eqnarray}  
\begin{eqnarray} 
e^{- i\lambda \, {a}^{k}} \, \Lambda^{i}_{\ 0}\,   e^{ i\lambda \, {a}^{k} }
 & = & { \Lambda^{i}_{ \ 0}   \over 
{\left(\Lambda^{0}_{\ 0} -1 \right) \over 2\kappa^{2} } \,  \lambda^{2} - 
{\Lambda^{k}_{\ 0} \over \kappa}\,  \lambda + 1  }  
\label{eluk22c} 
\end{eqnarray}  
\begin{eqnarray} 
e^{- i\lambda \, {a}^{k}} \, \Lambda^{0}_{\ k} \, e^{ i\lambda \, {a}^{k} } 
& = &   {    \Lambda^{0}_{ \ k}  + 
{\Lambda^{k}_{ \ k}\left(\Lambda^{0}_{ \ 0} - 1\right) + \Lambda^{0}_{ 
\ k} 
\Lambda^{k}_{ \ 0} \over \kappa} \, \lambda  \over 
{\left(\Lambda^{0}_{\ 0} -1 \right) \over 2\kappa^{2} }  \,  \lambda^{2}- 
{\Lambda^{k}_{\ 0} \over \kappa}\,  \lambda + 1  }  
\label{eluk22d} 
\end{eqnarray}  
\begin{eqnarray} 
e^{- i\lambda \, {a}^{k}} \, \Lambda^{k}_{\ k}\,  e^{ i\lambda \, {a}^{k} }
 & = &  {   -   {\left( 
\Lambda^{k}_{ \ k}\left(\Lambda^{0}_{ \ 0} - 1\right) + \Lambda^{0}_{ 
\ k}  
\Lambda^{k}_{ \ 0} \right)   \over  2 \kappa^{2}  }\, \lambda^{2} - 
{\Lambda^{0}_{ \ k} \lambda \over \kappa} + \Lambda^{k}_{\ k} \over  
{\left(\Lambda^{0}_{\ 0} -1 \right) \over 2\kappa^{2} } 
\,  \lambda^{2}  -  {\Lambda^{k}_{\ 0} \over \kappa}\,  \lambda + 1  } 
\label{eluk22e} 
\end{eqnarray}  
\begin{eqnarray} 
e^{- i\lambda \, {a}^{k}} \, \Lambda^{0}_{\ i}\,  e^{ i\lambda \, {a}^{k} }
 & = &  { \Lambda^{0}_{ \ i}+ 
{\left(\Lambda^{0}_{ \ 0} - 1\right) + \Lambda^{k}_{ \ i} + 
\Lambda^{k}_{ \ 0}  \Lambda^{0}_{ \ i} \over  \kappa  } \, \lambda 
 \over {\left(\Lambda^{0}_{\ 0} -1 \right) \over 2\kappa^{2} }
\,  \lambda^{2}  - 
{\Lambda^{k}_{\ 0} \over \kappa}\,  \lambda + 1  } 
\label{eluk22f} 
\end{eqnarray}  
\begin{eqnarray} 
e^{- i\lambda \, {a}^{k}} \, \Lambda^{k}_{\ i}\,  e^{ i\lambda \, {a}^{k} }
 & = &  { \Lambda^{k}_{ \ i}  - { \Lambda^{0}_{ \ i} \over \kappa }
\, \lambda  + { 
\left(\Lambda^{0}_{ \ 0} - 1\right) + \Lambda^{k}_{ \ i} + 
\Lambda^{k}_{ \ 0} \Lambda^{0}_{ \ i} \over  2\kappa^{2}  }
 \, \lambda^{2}  \over 
{\left(\Lambda^{0}_{\ 0} -1 \right) \over 2\kappa^{2} } 
\,  \lambda^{2} - {\Lambda^{k}_{\ 0} \over \kappa }\,  \lambda + 1 } 
\label{eluk22g} 
\end{eqnarray}  
\begin{eqnarray} 
&& i\neq k 
\cr 
&& e^{- i\lambda \, {a}^{k}} \, \Lambda^{i}_{\ k}\,  e^{ i\lambda \, 
{a}^{k} }=
\cr
&&
 =  
\Lambda^{i}_{ \ k} - {1\over \kappa} \int\limits^{\lambda }_{0} d\mu \, {
\Lambda^{i}_{ \ 0}\left[ -{  \Lambda^{k}_{ \ k} 
\left(\Lambda^{0}_{ \ 0} - 1\right) + \Lambda^{0}_{ \ k} \Lambda^{k}_{ 
\ 0}
\over 2 \kappa^{2}  } \, \mu^{2} - {\Lambda^{0}_{ \ k}
\over \kappa}\, \mu + \, \Lambda^{k}_{ \ k}  \right]
 \over   \left[ 
{\left(\Lambda^{0}_{\ 0} -1 \right) \over 2\kappa^{2} } 
\,  \mu^{2} -  {\Lambda^{k}_{\ 0} \over \kappa}\,  \mu + 1 
\right]^{2} } 
\label{eluk22h} 
\end{eqnarray}  
\begin{eqnarray} 
&& i\neq k, j\neq k 
\cr  
&& e^{- i\lambda \, {a}^{k}} \, \Lambda^{i}_{\ j}\,  e^{ i\lambda \, 
{a}^{k} }=
\cr
&&
 =  \Lambda^{i}_{ \ j}  - {1\over \kappa}
\int\limits^{\lambda }_{0} d\mu \, { \Lambda^{i}_{ \ 0} \left[ -  { 
\left(\Lambda^{0}_{ \ 0} - 1\right) \Lambda^{k}_{\ j} 
+ \Lambda^{k}_{ \ 0} \, \Lambda^{0}_{\ j} \over 2 \kappa^{2}  }
 \, \mu^{2} - {\Lambda^{0}_{ \ k}\over \kappa}\, \mu + \,
 \Lambda^{k}_{ \ j} \right] \over \left[ 
{\left(\Lambda^{0}_{\ 0} -1 \right) \over 2\kappa^{2} } 
\,  \mu^{2} - {\Lambda^{k}_{\ 0} \over \kappa}\,  \mu + 1 
\right]^{2} }  
\label{eluk22i}   
\end{eqnarray}    
\end{subequations}
In order to represent the  relation (\ref{eluk33}) in star product
framework we reproduce the multiplication (\ref{eluk33}) by a new
star product of the basic functions on classical Poincar\'{e}
group parameters:
\begin{eqnarray}\label{eluk37} 
&&  e^{ i( \alpha_{\kappa} v^{\kappa} +
  b^{\nu}_{\mu} \Lambda^{\mu}_{\ \nu})}
  \circledast
   e^{ i( \alpha^{\prime}_{\kappa} v^{\kappa} +
  b^{\nu \prime }_{\mu}  \Lambda^{\mu }_{\ \nu})}
\cr\cr  && \qquad  = e^{i(\Delta_{\mu}(\alpha, \alpha^{\prime})
v^{\kappa} + b^{\nu}_{\mu} f^{\kappa}_{\nu}
(g^{\rho}_{\sigma} (1, \vec{\alpha}^{\prime} ), \alpha_{0} )
+ b^{\nu \prime}_{\mu} \Lambda^{\kappa}_{\ \nu} )}
\end{eqnarray}
As it is seen from (\ref{eluk22a} --\ref{eluk22i} ) 
 the function                           
$g^{\rho}_{\sigma}$ are not linear in
$\widehat{\Lambda}^{\kappa}_{\ \nu}$, due to the quadratic 
commutator
(\ref{eluk21b}). On the other hand, due to   the commutativity
 (\ref{eluk21c}) the  formulae for $f^{\kappa}_{\nu}$ and
  $g^{\kappa}_{\nu}$ can be obtained in explicite form.

\section{Final Remarks}
                                      
We  would like to make the following comments:

i) It should be observed that the nice coproduct formula
 (\ref{eluk210}) for noncommutative translations can not be
 extended to the Lorentz sector. One can pose the question 
whether
 by a suitable choice of \underline{noncommutative}
 $b^{\mu}_{\ \nu}$ such extension can be achieved.

 ii) The relations (\ref{eluk21a}--\ref{eluk21c}) describe a
 quadratic algebra, which implies that in the exponential on rhs
  of (\ref{eluk37}) there are arbitrary powers of
  $\Lambda^{\kappa}_{\ \nu}$. It should be  observed, however, that
   the multiplication formula (\ref{eluk37}) has an explicite
   form.

   iii) Using the star  product (\ref{eluk37}) one can discuss the
   covariance of $\kappa$--deformed local field theory under the
   $\kappa$--deformed relativistic transformations. At present it
   is only clear how to show in the star--product framework the
   covariance under the subgroup of noncommutative translations
   (see also \cite{rluk14}).

\section*{Acknowledgment}
  One of the authors (JL) would like to thank the members of 
Departamento 
de Fisica, Universidad de Valencia for the hospitality and 
Generalitat 
Valenciana for financial support.

\end{document}